\documentclass[10pt,letterpaper,twocolumn]{article} 

\usepackage{ol2}
\usepackage[draft]{hyperref}
\usepackage{amsmath}

\begin{document}

\twocolumn[ 

\title{Stable single-photon interference in a 1 km fiber-optical Mach-Zehnder interferometer with continuous phase adjustment}


\author{G. B. Xavier,$^{1*}$ and J. P.  von der Weid,$^1$}

\address{
$^1$Center for Telecommunication Studies, Pontifical Catholic University of Rio de Janeiro, R. Marqu\^es de S\~ao Vicente, 225 G\'avea - Rio de Janeiro - Brazil - 22451-900\\
$^*$Corresponding author: guix@opto.cetuc.puc-rio.br
}

\begin{abstract}We experimentally demonstrate stable and user-adjustable single-photon interference in a 1 km long fiber-optical Mach-Zehnder interferometer, using an active phase control system with the feedback provided by a classical laser. We are able to continuously tune the single-photon phase difference between the interferometer arms using a phase modulator, which is synchronized with the gate window of the single-photon detectors. The phase control system employs a piezoelectric fiber stretcher to stabilize the phase drift in the interferometer. A single-photon net visibility of 0.97 is obtained, yielding future possibilities for experimental realizations of quantum repeaters in optical fibers, and violation of Bell's inequalities using genuine energy-time entanglement.\end{abstract}


 ] 

\noindent Research in the distribution of single-photons over long distances in optical fibers with high-fidelity is of great interest, with practical applications in quantum communication \cite{Gisin_Nat_Photon} and in more fundamental physics such as tests of Bell's inequalities \cite{Zeilinger_PRL_1999}. More recently the need for long-distance distribution of single-photons in an interferometric configuration has arisen, focusing in some quantum repeater protocols \cite{Geneva_repeater}, quantum key distribution \cite{Noh_PRL} and fundamental tests of Bell's inequalities \cite{Cabello_PRL}.     

The interference of single-photons is routinely performed on optical tables, however, it is still a challenge in long-distance fiber optical interferometers. All interferometers are sensitive to path-length fluctuations leading to a change in the optical intensity as a function of time at the output of the interferometer. It is well known that $I\propto 1+ \text{cos} [\Delta\phi]$, where $I$ is the optical intensity at one of the output ports of the interferometer and $\Delta\phi$ is the phase difference between the two arms \cite{Born_Wolf}. In the case of long fiber optical interferometers we should emphasize that $\Delta\phi$ is heavily dependent on the environment and its rate of fluctuation can be in the kHz range \cite{Minar_PRA}. Nevertheless, in spite of these difficulties, stable classical and single-photon interferences have been demonstrated in long fiber-optical interferometers \cite{Delage_Opex, Noh_Opex}. 

In this letter we demonstrate for the first time, to the best of our knowledge, stable single-photon interference in a long interferometer in a fiber optical Mach-Zehnder (MZ) configuration. Such an interferometer is of great interest for quantum repeaters \cite{Geneva_repeater} and violation of Bell's inequalities employing genuine energy-time entanglement \cite{Cabello_PRL}. For the test of Bell's inequalities, a controllable phase shift between the interferometer arms is necessary to perform Alice and Bob's measurement choices \cite{Glima_PRA}, which we are able to demonstrate in our experiment, using a phase modulator synchronized with the single-photon detectors. Our scheme is also an improvement over \cite{Noh_Opex} since we are able to provide a true continuous tunable phase-shift. 

The experimental setup is shown in Fig. 1. One external cavity tunable laser operating in continuous wave (CW) mode, set at $\lambda_{\text{PH}}  = 1547.72$ nm, is used to provide feedback to the active phase stabilization system. Its output spectrum is filtered using an optical circulator together with a fiber Bragg grating (FBG) designed to reflect $\lambda_{\text{PH}}$ in order to remove the amplified spontaneous emission from the laser, which would otherwise fall in-band within the single-photons wavelength. The single-photons themselves are generated by an attenuated distributed feedback (DFB) laser also working in CW mode, with center wavelength $\lambda_{\text{Q}}  = 1546.12$ nm. The optical attenuator is adjusted for an average photon number at the input of the interferometer of 0.1 photon per detection window.  Both lasers occupy a channel (henceforth referred to as the classical and quantum channels) in the 100 GHz dense wavelength division multiplexing (DWDM) grid, such that they are located 200 GHz apart. They are both combined in a DWDM multiplexer with 1.6 dB insertion loss, and $>$ 40 dB isolation between adjacent 100 GHz channels. The combined filtering between the FBG and DWDM provides $>$ 100 dB isolation between the classical and quantum channels, which is enough to avoid cross-talk based noise from reaching the single-photon detectors.

\begin{figure}[htb]
\centerline{\includegraphics[width=7.5cm]{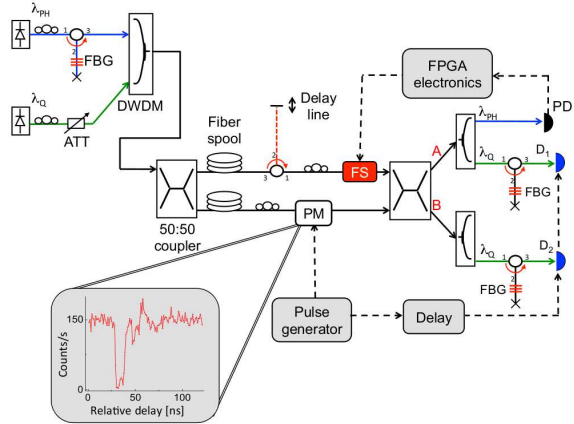}}
\caption{(Color online) Experimental setup. Solid lines represent optical fibers and connections, dashed black lines stand for electrical cables and red dashed depicts free-space.  ATT: Variable optical attenuator; D$_1$ and D$_2$: Single-photon counting modules; DWDM: Dense wavelength division multiplexer; FBG: Fiber Bragg grating; FPGA: Field programmable gate array; FS: Piezoelectric fiber stretcher; PD: Photodetector; PM: Electro-optic phase modulator. Inset shows the counts/s at D$_1$ when changing the relative delay from the pulse generator to the single-photon detector.}
\end{figure}

The MZ interferometer is composed of two 50:50 fiber couplers and two single-mode optical fibers in a 1 km spool. Both fibers are inside the same rugged commercial cable, composed of 6 single-mode and 6 multi-mode fibers. The thick cover provided by the cable helps to shield the interferometer from acoustic, mechanical and thermally induced vibrations, and it is more similar to a real installed fiber cable when compared to a typical laboratory fiber optical spool with just the primary coating applied. After the second 50:50 coupler, a DWDM demultiplexer is placed at output port A and splits both wavelengths. The optical intensity of the control signal $\lambda_{\text{PH}}$ is monitored with photodetector PD, which has a bandwidth of 1 MHz. The single-photons ($\lambda_{\text{Q}}$) exit the DWDM and pass through another circulator + FBG, but this time centered at $\lambda_{\text{Q}}$, whose purpose is to filter out photons leaked by cross-talk in the demultiplexing procedure. At the other output port of the interferometer (B) an exact identical filtering combination is used for the single-photons. The total filtering bandwidth in the quantum channel is $\sim$ 50 GHz. Raman spontaneous scattering is not a limiting factor in this experiment due to the 1 km length of the optical fiber and because of the low - 17.0 dBm launch power of the classical control channel \cite{Guix_EL_Raman}.

The interferometer is phase-stabilized by maintaining the optical intensity of the classical feedback signal at PD constant and in quadrature \cite{Noh_Opex}, using a commercial piezoelectric fiber stretcher. The electrical feedback detected signal is processed using field programmable gate array (FPGA) electronics with a proportional-integral-derivative (PID) algorithm, and is then amplified by a custom home made electronic driver, and connected to the stretcher, thus closing the feedback loop. The highest natural frequency component of the optical intensity fluctuations at the output of the interferometer, as measured with an oscilloscope, is in the order of 100 Hz. Our control system has a maximum frequency response, as limited by the fiber stretcher, of $\sim$ 5 kHz, which is sufficient to phase-lock the 1 km interferometer. 

The single-photons are detected by two InGaAs single-photon counting modules D$_1$ and D$_2$, working under gated mode \cite{Gisin_RMP}, with 2.5 ns gate window, an overall efficiency of 15\% and a measured dark count probability per gate of $9.33\times 10^{-6}$ and $4.14\times 10^{-5}$ respectively, at 166 kHz repetition rate. The length mismatch of the interferometer is adjusted using a free-space delay line combined with an optical circulator, with a total incurred loss of 3 dB. For this balancing procedure a low coherence light source replaces the classical feedback signal
at the interferometer input, and the filtered interference fringes are measured at PD, balancing the arms within 1 mm, which is much less than the coherence lengths of the single-photons and the classical feedback signal.

In order to provide an adjustable phase difference between the two arms, we perform the following: we write that the optical intensity at the output of the interferometer is $I(t,V) \propto 1+\text{cos}[\Delta\phi(t)+\Delta\varphi(V)]$, where $\Delta\phi(t)$ are the phase fluctuations caused by the environment and $\Delta\varphi(V)$ is the adjustable phase conditioned on a parameter $V$. Since the control system is able to keep the interferometer phase stabilized, the optical intensity is independent of $\Delta\phi$ and of time. By implementing a phase shift that does not disturb the control system, this shift can be made to be adjustable. This is done through a telecom electro-optic phase modulator (PM) in one of the interferometer arms. A generator creates 10 ns wide electrical pulses that drives the PM, therefore causing a phase difference between the interferometer arms in a time scale much faster than the natural oscillation of the interferometer and the frequency response of the phase stabilization system. Therefore with the phase control active we have $I(V) \propto 1+\text{cos}[\Delta\varphi(V)]$ where $V$ is the voltage applied to the PM. 

Clearly we need to synchronize the detection system with the imposed phase difference pulse $\Delta\varphi$. This is quite straightforward in many quantum communication experiments, when working with gated detectors, such as in our experiment. An electronic delay generator (DG) is triggered by the pulse generator, and two of the DG's outputs are connected to D$_1$ and D$_2$. The inset of Fig. 1 shows the shape of a typical modulation pulse as detected at D$_1$ when applying maximum driving voltage to the PM and changing the relative delay between the pulse generator and the detector. The observed ``ringing'' effect after the main pulse is due to a slight impedance mismatching in the phase modulator, and it does not affect the experiment, since the 2.5 ns detection window is placed in the beginning of the pulse. The total delay needed to synchronize the modulation pulse with the detectors is $\sim$ 5.8 $\mu \text{s}$, which limits the repetition rate that the system can work on to $\sim$ 166 kHz. 

A manual polarization controller is placed before the PM in order to obtain maximum depth of modulation, while a second one, placed in the other arm, is used to obtain maximum visibility of the interference fringes. We observed no significant polarization drift during a time scale of several minutes. We also verified with a polarimeter that the driving voltage applied to the fiber stretcher does not change the output state of polarization (SOP) significantly.  Another manual polarization controller is placed at the output of each laser in order to have identical input SOPs for both channels, in order to facilitate alignment of the experiment. We first adjust the SOP overlap at the second 50:50 coupler to be near unity, and the PM to obtain maximum probability to detect single-photons at one of the detectors ($\Delta\varphi = 0$ or $\pi$). We then record the number of counts per second from both detectors and plot the raw results in Fig. 2. After approximately 250 s the phase stabilization is turned off, demonstrating its effectiveness. The average net visibility (subtracting the dark counts of the detectors), which is defined as $V=|(C_2 - C_1)/(C_2+C_1)|$ where $C_1$ and $C_2$ are the counts per unit of time in detectors D$_1$ and D$_2$ respectively, for the period of time until the phase control is turned off is $0.971 \pm 0.021$. The increased observed dispersion in the curve for D$_2$ is mainly due to the statistical fluctuation of the photon detection process. After the phase control is turned off, the lower contrast in the single-photon interference fringes is due to the 1 s integration time of the detectors.

\begin{figure}[htb]
\centerline{\includegraphics[width=7.5cm]{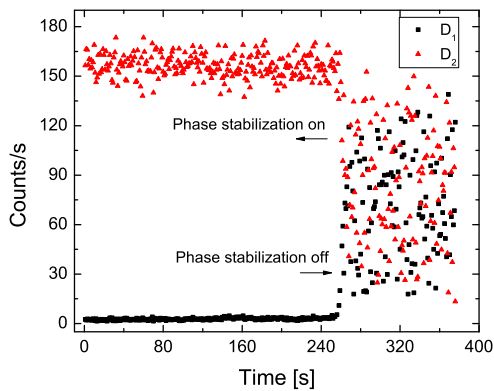}}
\caption{(Color online) Detected counts per second in both D$_1$ and D$_2$ with the phase control active and the system aligned. At around 250 s the phase control is turned off.}
\end{figure}

In Fig. 3 we plot the counts per second as a function of the driving voltage in the PM, up to a maximum possible voltage of 6.8 V. Each measured point has an integration time of 10 s, with the results displayed as the average number of counts per second. We clearly observe that it is possible to perform continuous tuning of the phase difference between the two arms using the phase modulator. 

We have shown stable and adjustable single-photon interference with a 1 km long MZ fiber-optical interferometer, obtaining constant 0.97 net visibility over several tens of seconds. Longer stability is possible by providing better thermal insulation to the components placed outside of the fiber spool, using components with shorter pigtails as well as employing a faster actuator, creating possibilities for a real field trial. To the best of our knowledge this is the first time that user-adjustable single-photon interference is obtainable in such a long distance interferometer. The same level of adjustment in the interference is possible with classical light, as long as  synchronization is provided between the phase modulator and the detector. Our results open up new experimental possibilities in quantum communication requiring phase stabilization in long distance MZ interferometers, such as quantum repeaters and fundamental tests of Bell's inequalities using genuine energy-time entanglement.

We acknowledge technical assistance with the electronics from T. R. da Silva, and financial support from FAPERJ, CAPES and CNPq. 

\begin{figure}[htb]
\centerline{\includegraphics[width=7.5cm]{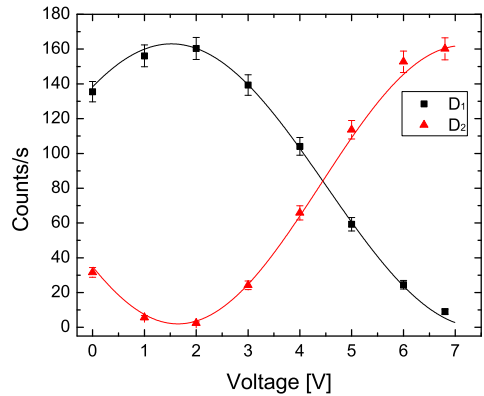}}
\caption{(Color online) Photon detections at D$_1$ and D$_2$ as a function of the phase modulator driving voltage. Solid lines demonstrates the theoretical fit to the data. Error bars represent the statistical fluctuations of the detection process.}
\end{figure}



\begin{thebibliography}{99}


\bibitem{Gisin_Nat_Photon} 
N. Gisin and R. Thew, Nat. Photon. \textbf{1}, 165 (2007).

\bibitem{Zeilinger_PRL_1999} 
G. Weihs, T. Jennewein, C. Simon, H. Weinfurter and A. Zeilinger, Phys. Rev. Lett. \textbf{81}, 5039 (1998).

\bibitem{Geneva_repeater} 
N. Sangouard, C. Simon, H. de Riedmatten and N. Gisin, Rev. Mod. Phys. \textbf{83}, 33 (2011).

\bibitem{Noh_PRL} 
T.-G. Noh, Phys. Rev. Lett. \textbf{103}, 230501 (2009). 

\bibitem {Cabello_PRL} 
A. Cabello, A. Rossi, G. Vallone, F. De Martini and P. Mataloni, Phys. Rev. Lett. \textbf{102}, 040401 (2009).

\bibitem{Born_Wolf} 
M. Born and E. Wolf, \textit{Principles of Optics, 7th (expanded) edition} (Cambridge University Press, 1999).

\bibitem{Minar_PRA} 
J. Min\'a\v{r}, H. de Riedmatten, C. Simon, H. Zbinden and N. Gisin, Phys. Rev. A. \textbf{77}, 052325 (2008).

\bibitem{Glima_PRA}
G. Lima, G. Vallone, A. Chiuri, A. Cabello and P. Mataloni, Phys. Rev. A \textbf{81}, 040101(R) (2010).

\bibitem{Delage_Opex} 
L. Delage and F. Reynaud, Opt. Express \textbf{9}, 267  (2001).

\bibitem{Noh_Opex}
S.-B. Cho and T.-G. Noh, Opt. Express \textbf{17}, 19027 (2009).

\bibitem{Guix_EL_Raman}
G. B. Xavier and J. P. von der Weid, Electron. Lett. \textbf{46}, 1071 (2010).

\bibitem{Gisin_RMP}
N. Gisin, G. Ribordy, W. Tittel and H. Zbinden , 
Rev. Mod. Phys. \textbf{74}, 145 (2002).


\end{thebibliography}
\end{document}